\documentstyle[editedvolume,epsfig]{crckapb}

\newcommand{\gsim}{\raisebox{-0.3ex}{\mbox{$\stackrel{>}{_\sim} \,$}}}

\begin{opening}
\title{Kinematics of Radio Pulsars}

\author{R. Ramachandran}
\institute{Netherlands Foundation for Research in Astronomy \\
           Postbus 2, 7990 AA Dwingeloo, The Netherlands.}

\end{opening}

\runningtitle{Kinematics of radio pulsars}

\begin{document}


\section{Introduction}
A very good fraction of stars in the sky are believed to be in binary (or
multiple) systems. Therefore, we expect a good fraction of pulsars to have
originated in binary systems. However, observations of pulsars have shown that
most of the pulsars are not in binary system now. This would imply that
these pulsars must have become solitary after the disruption of their progenitor
binary system during (or shortly after) the supernova explosion. 

Well before the discovery of pulsars, the work Blaauw (1961) showed that if a
star explodes in a binary system, if the mass lost in that process is greater
than half the initial total mass, then the final system will be unbound. This
would imply that the resultant pulsar will have a spatial velocity equal to the
orbital velocity of the exploded star at the time of the explosion.

After the discovery of pulsars, from their statistical analysis of the pulsar
population, Trimble and Rees (1970) showed that in order to explain the
distribution of pulsars, one needs a large spatial speeds for pulsars when
compared to what they would have got from the orbital velocities of their
progenitor binary systems. 

Following this, two main mechanisms were suggested to explain the origin of
pulsar velocities (other than from the binary orbital velocities). 

\begin{enumerate}
\item Shklowskii (1970) conjectured that supernova explosions are asymmetric in
      nature, and that helps pulsars to acquire substantial velocities during
      their birth.
\item In 1975, Harrison \& Tademaru came up with another mechanism, where
      pulsars acquire their velocities by ``Rocket-effect'', by having
      Oblique-offcentred magnetic dipole. 
\end{enumerate}

They showed that if we have such a magnetic field arrangement the radiation
pattern of that dipole is asymmetric with respect to the two poles defined by
the rotation axis. The reaction force on the neutron star in such a case is
prportional to the fifth power of the angular velocity ($\Omega^5$). This
implies two important results: ($a$) Young new-born fast-rotating pulsars can be
accelerated to very high velocities, ($b$) This effect is strong enough to
disrupt the binary system where the pulsars were born! Therefore, with this
mechanism, the authors could explain both the problems (that pulsars have large
spatial velocities, and they are solitary).

Although this method, in principle could give qualitative explanation to the
nature of pulsar population, it failed a very important observational test. As
Morris, Radhakrishnan \& Shukre (1976) showed, the acceleration of a pulsar in
the direction of its rotation axis would necessarily mean that in the plane of
the sky the projected direction of rotation axis is the same as that of the
measured proper motion direction. From the measured proper motion and
polarisation data, they proved that this correlation is not seen (see also
Anderson \& Lyne 1983; Lorimer et al. 1995; Deshpande, Ramachandran,
Radhakrishnan 1999). Therefore, we can assume that this elegant mechanism is not
relevant for understanding the origin of pulsar velocities.

In this paper, I would summarise the main developments in this subject so
far. 

\begin{figure}
\epsfig{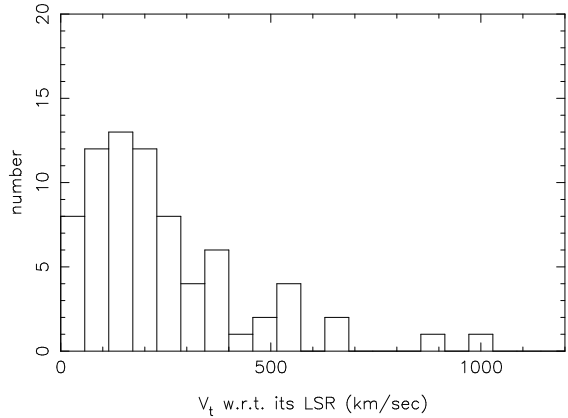}
\caption{Distribution of observed transverse velocities after correcting for
differential Galactic rotation.}
\label{fig:obsvelo}
\end{figure}

\begin{table}
\begin{center}
\begin{tabular}{lcccccccc}\hline
{\bf Jname} & {\bf l} & {\bf b} & {\bf d} & {\bf $\mu_{\alpha}$} & {\bf
err} & {\bf $\mu_{\delta}$} & {\bf err} & $V_{t}$ \\
  & (deg) & (deg) & (kpc) & (mas/yr) &  & (mas/yr) &  & km/s \\ \hline\hline
0152--1637 &   179.3 & --72.5 &    .79   &  --10 &   50 & --150 &   50 &  563 \\
0452--1759 &   217.1 & --34.1 &  $>$3.14 &   19  &    8 &   35  &   18 &  $>$593 \\
0525+1115  &   192.7 & --13.2 &  $>$7.68 &   30  &    7 &   --4 &    5 & $>$1102 \\
0601--0527 &   212.2 & --13.5 &  $>$7.54 &   18  &    8 &  --16 &    7 &  $>$861 \\
0738--4042 &   254.2 &  --9.2 & $>$11.03 &  --56 &    9 &   46  &    8 & $>$3789 \\
0826+2637  &   197.0 &  31.7  &    .38   &   61  &    3 &  --90 &    2 &  644 \\
0922+0638  &   225.4 &  36.4  &  $>$2.97 &   13  &   29 &   64  &   37 &  $>$919 \\
1509+5531  &    91.3 &  52.3  &   1.93   &  --73 &    4 &  --68 &    3 &  913 \\
1604--4909 &   332.2 &   2.4  &   3.59   &  --30 &    7 &   --1 &    3 &  511 \\
1645--0317 &    14.1 &  26.1  &   2.90   &   41  &   17 &  --25 &   11 &  660 \\
1709--1640 &     5.8 &  13.7  &   1.27   &   75  &   20 &  147  &   50 &  993 \\
1720--0212 &    20.1 &  18.9  &  $>$5.43 &   26  &    9 &  --13 &    6 &  $>$748 \\
1935+1616  &    52.4 &  --2.1 &   7.94   &    2  &    3 &  --25 &    5 &  944 \\
2149+6329  &   104.3 &   7.4  & $>$13.64 &   14  &    3 &   10  &    4 & $>$1112 \\
2225+6535  &   108.6 &   6.8  &   1.95   &  144  &    3 &  112  &    3 & 1686 \\
2305+3100  &    97.7 & --26.7 &  $>$3.93 &   13  &    8 &  --33 &    6 &  $>$661 \\ \hline
\end{tabular}
\caption[]{Pulsars with possible transverse velocities greater than 500
km/sec. Columns 2 \& 3 give the galactic longitude and latitude, col. 4 gives
the distance calculated using Taylor \& Cordes (1993) model, 5--8 give the
measured proper motions along RA and Dec and their errors, and the last column
gives the transverse velocity.}
\label{tab:table1}
\end{center}
\end{table}

\section{Observational facts}
There are two `direct' ways of measuring pulsar proper motions. The first is by
an interferometer, where observations at two well separated epochs can give an
accurate measurement of positional displacement of the pulsar in the sky. The
second is by Timing measurements, where the positional errors (due to the proper
motion) lead to annual oscillations in timing residuals, and this gives an
estimate of their proper motion. So far, we have proper motion measurements for
about 100 pulsars by the above two methods.

Figure \ref{fig:obsvelo} shows the observed transverse velocity of pulsars after
correcting for the differential galactic rotation. As we can see, though the
maximum velocities go all the way to about 1000 km/sec, majority of pulsars have
velocities of about 150 to 200 km/sec. While understanding this distribution,
apart from the measurement errors associated with these proper motion
measurements, it is also important to appreciate the uncertainties in the
estimated distances to these pulsars ($V_{t} = \mu\times D$, where $\mu$ and $D$
are the measured proper motion and distance to the pulsar). Table
\ref{tab:table1} gives a list of pulsars for which transverse velocities are
suspected to be greater than 500 km/sec. Distances to these objects are
estimated with the help of the Galactic free-electron density distribution model
of Taylor \& Cordes (1993). It is clear that for half of them, we have only a
lower limit on their distance. Even for these pulsars, a ``lower limit'' doesn't
necessarily mean that the actual distance is greater than the lower limit! As
Deshpande \& Ramachandran (1998) showed from their scattering measurements, the
distance to PSR J0738--4042 is as small as about 4 kpc, whereas the lower limit
from the Taylor \& Cordes model is $>11$ kpc. After accounting for all these
uncertainties, what we can conclude from this table is that there are a few
pulsars, like 0826+2637, 1509+5531, and 2225+6535, which have sufficiently well
measured proper motions and well determined distances, which are moving with
transverse velocities well in excess of 500 km/sec. Therefore, we do see some
fraction of pulsars having such high velocities.

Observationally, many reasons have emerged over the years to justify the idea
that one needs an impulse (`kick') due to asymmetric supernova explosion. They
are,
\begin{itemize}
\item The rotation axis of the Be-star companion of PSR J0045--7319 is
      misaligned with the orbital angular momentum axis, strongly suggesting
      that the supernova explosion must have been asymmetric (Kaspi et al. 1996;
      Lai 1996).
\item Large orbital eccentricities of Be X-ray binaries
\item Low incidence of double neutron star binary systems in the
      Galaxy. Observational estimates show a birthrate of $\le 10^{-5}$
      yr$^{-1}$ (Phinney 1991; Narayan et al. 1991; van den Heuvel \& Lorimer
      1996). To reproduce this, we need a kick speed of a few hundred km/sec
      (Portegies Zwart \& Spreew 1996; Ramachandran 1996; Lipunov et al. 1996;
      Bagot 1996; Fryer \& Kalogera 1997).
\item Low incidence of Low Mass X-ray Binaries, and their kinematics in the
      Galaxy (van Paradijs \& White 1995; Brandt \& Podsiadlowski 1995;
      Ramachandran \& Bhattacharya 1997; Cordes \& Chernov 1997; Kalogera 1996;
      1998; Tauris \& Bailes 1996).
\item It is necessary to have kick velocities to produce systems like PSR
      B1913+16 (Wex et al. 2000)
\end{itemize}

Considering all the above observational constraints, it seems that an asymmetric
supernova explosion leading to a ``kick'' seems to be most probable. However, it
is also important to understand the fractional contribution of binary orbital
velocities to the observed pulsar velocities.

\section{Distribution of pulsar speeds}
\label{sec-selec}
To understand the intrinsic distribution of pulsar speeds in the Galaxy, we need
to do a detailed statistical analysis of the pulsar population. There are many
observational selection effects biasing our sample of pulsars, and they need to
be understood and corrected for, before attempting to quantify any statistical
property of the pulsar population. A brief description of the selection effects
is as follows (see Lorimer's talk in this volume for further details).
\begin{itemize}
\item Pulsar radio luminosities have a wide ranging distribution. Faint objects
      don't get detected, biasing our sample against the `fainter' side of the
      luminosity distribution.
\item The probability of detecting farther pulsars reduces due to the smearing
      of the pulse due to dispersion and scattering of radio signals in the
      interstellar medium.
\item Slow pulsars don't migrate to large heights above the plane, whereas fast
      pulsars do. This gives an over-estimation of the scale factors for slow
      pulsars (Hansen \& Phinney 1997). 
\item Farther slow pulsars don't show significant (measurable) proper motion,
      biasing the sample against slow pulsars.
\item Distances are over-estimated due to the presence of (unaccounted) HII
      regions (or Str\"{o}mgren spheres of OB stars) which introduce significant
      extra dispersion measure. This results in over-estimation of the
      transverse velocities (Deshpande \& Ramachandran 1998).
\end{itemize}
Any statistical study of pulsar population must take into account all these
effects to get an unbiased idea about the properties of the pulsar population.

\begin{figure}
\epsfig{file=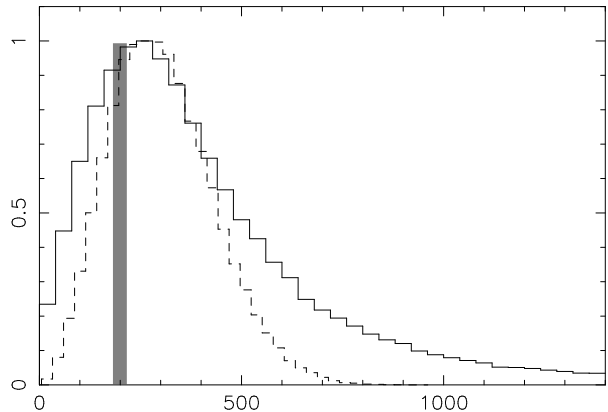,width=5.5cm}
\caption[]{Derived distribution of three dimensional speeds of pulsars. Solid
line is by Lyne \& Lorimer (1994), `Dash' line, by Hansen \& Phinney (1997), and
the shaded distribution by Blaauw \& Ramachandran (1998).}
\label{fig:distrib}
\end{figure}

Figure \ref{fig:distrib} shows the distribution of three dimensional speeds of
the pulsar population estimated by three different works.  Lyne \& Lorimer
(1994) considered a sample of 28 pulsars whose ages are less than 3 Myr for
their analysis. With the assumption that there are no selection effects on such
sample of young pulsars, they estimated the speed distribution, which has a long
`tail' extending to well above 1000 km/sec. This predicts an average speed of
about 450 km/sec, with an r.m.s. speed of 535 km/sec and the peak of the
distribution around 250 km/sec (solid line in the figure). Hansen \& Phinney
(1997), from their detailed analysis of the Galactic population of pulsars came
to the conclusion that the pulsar population is quite consistent with a
Maxwellian distribution with an one-dimensional r.m.s. velocity of about 190
km/sec (`dash' curve). With their detailed analysis of the local population of
pulsars, Blaauw \& Ramachandran (1998) concluded that the speed distribution of
pulsars in the solar neighbourhood is consistent with even a `delta-function'
distribution at about 200 km/sec! The idea that predominant fraction of pulsars
may be moving with substantially lower velocities ($\sim 200-250$ km/sec) was
also supported by the analysis by Hartman (1997). Therefore from all these
analyses, we can conclude that bulk of the population is moving with 3-D speeds
of about 200 -- 250 km/sec.

\section{Kinematics of pulsars in the Galaxy}
From the above analysis, it is clear that pulsars do have large peculiar speeds
of the order of about 200 -- 250 km/sec. With the addition of the Galactic
rotation, the resultant velocities of these objects could be very much
comparable to the escape velocity of the Galaxy. Therefore, study of the
evolution of these objects in the Galactic potential becomes important. I will
present here a `quick' Monte Carlo simulation to understand some important
aspects of their evolution. We will address some of the questions related to
their kinematical properties like `asymmetric drift' in the Galactic plane,
their migration along the Galactocentric radius, etc. Let us generate a large
number of pulsars in the Galaxy with the following properties.
\begin{itemize}
\item The surface density distribution of these objects along the Galactocentric
      radius is a Gaussian, with an r.m.s. of $\sigma_R = 4.5$ kpc. 
\item Along the height from the galactic plane, let us assume a scale-height of
      75 pc. 
\item Evolve each pulsar after including the Galactic rotation contribution
      corresponding to its place of birth, for a total length of time $T$, where
      $T$ is distributed uniformly between zero and a maximum of $2\times 10^9$
      yrs. After evolving each object, store the values of its initial   
      and final coordinates, and its final velocity components in the Galaxy. 
\item Study the spatial and velocity distributions, after correcting for the
      observational selection effects as indicated in section \ref{sec-selec}.
\end{itemize}

The Galactic potential function is assumed to be the one given by Kuijken \&
Gilmore (1989):
\begin{equation}
\Phi(R,z)\;=\;\frac{-M}{\left[\left(a + \sqrt{z^2+h^2}\right)^2 + b^2 +
R^2\right]^{1/2}}
\end{equation}
\noindent
This potential function has three components, namely the Disc-Halo, Nucleus, and
the Central Bulge. For the Disc-Halo component, $\sqrt{z^2+h^2} = \sum^3_{i=1}
\beta_i \sqrt{z^2+h_i^2}$. The values of all the constants are tabulated in
Table \ref{tab:table2}

\begin{table}
\begin{center}
\begin{tabular}{llll}\hline
{\bf Parameter} & {\bf Disc-Halo} & {\bf Nucleus} & {\bf Bulge} \\ \hline\hline
{\bf Mass} (M$_{\odot}$) & $1.45\times 10^{11}$ & $9.3\times 10^{11}$ & $1\times
   10^{10}$ \\
{\bf $\beta_1$} & 0.4    &      &     \\
{\bf $\beta_2$} & 0.5    &      &     \\
{\bf $\beta_3$} & 0.1    &      &     \\
{\bf $h_1$}     & 0.325  &      &     \\
{\bf $h_2$}     & 0.090  &      &     \\
{\bf $h_3$}     & 0.125  &      &     \\
{\bf $a$}       & 2.4    &      &     \\
{\bf $b$}       & 5.5    & 0.25 & 1.5 \\ \hline
\end{tabular}
\caption[]{Parameters of the Galactic potential function by Kuijken \& Gilmore
(1989).}
\label{tab:table2}
\end{center}
\end{table}

\subsection{Asymmetric Drift}
When a pulsar evolves in the galaxy, due to its peculiar velocity it migrates
along both $z$ and $R$. When a pulsar is born in the Galaxy, its initial angular
momentum is determined by the Galactocentric radius at which the pulsar is born,
and its velocity around the galactic centre. Given this, and the flat rotation
curve of the Galaxy, when the pulsar migrates to a different galactocentric
radius, it either leads ahead, or lags behind the local flow, depending on
whether the present galactocentric radius is less or more than its initial
radius. Since, on the average, objects flow to the outer portions of the Galaxy
from the inner portions, we expect more objects laging behind. In other words,
any virialised population with a significant radial velocity component will
revolve more slowly around the Galactic centre. This effect is known as
`asymmetric drift', and is seen in many stellar populations in the Galaxy
(Mihalas \& Binny 1981).

Pulsars, though they are relatively young objects, they have far greater
velocity than almost all the stellar populations. Therefore, even within a
relatively short time ($\sim 10$ Myr) they exhibit this important
property. Hansen \& Phinney (1997) showed through their detailed analysis that
90\% of those pulsars whose ages are greater than about 4 Myr show this effect.

\begin{figure}
\epsfig{file=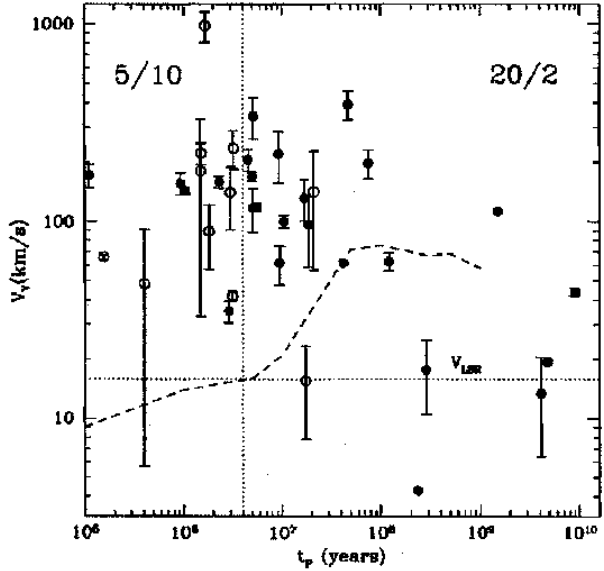,width=8.5cm}
\caption[]{Plot of characteristic age Vs azimuthal velocity ($V_y$) in the Galaxy
(Hansen \& Phinney 1997). Solid circles indicate positive $V_y$. About 90\% of
the pulsars with ages greater than 4 Myr seem to have positive $V_y$, strongly
indicating asymmetric drift. The `dash' line indicates the evolution of the mean
asymmetric drift velocity from their calculations.}
\label{fig:drift}
\end{figure}

Figure \ref{fig:drift} shows the plot of Hansen \& Phinney of the characteristic
age and the $V$ component (velocity in the azimuthal direction with respect to
the Galactic centre) derived from their proper motion
measurements. Though they have calculated only the azimuthal velocity with
respect to the position of the Sun (not with respect to the position of each of
the pulsars), it gives a clear idea about the significance of the asymmetric
drift.

\begin{figure}
\epsfig{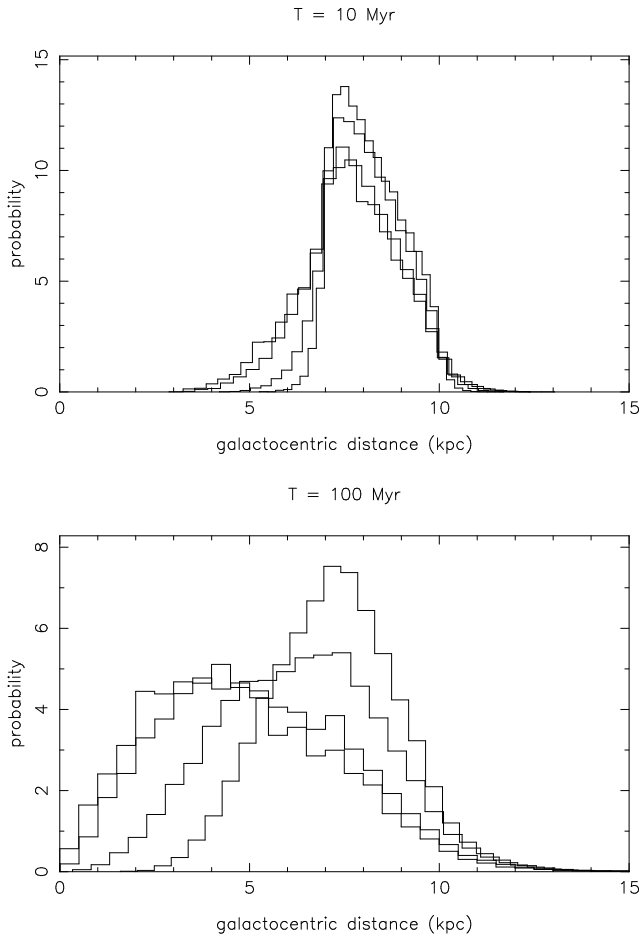}
\caption[]{Distribution of the initial galactocentric radius for those objects
which end up in the Solar neighbourhood after their evolution. The two panels
correspond to a maximum evolution time of 10 Myr and 100 Myr, respectively. The
distributions (from the narrowest to the broadest) correspond to the assumed
initial Maxwellian speed distributions with 1--D r.m.s. of 50, 100, 250, and 500
km/sec, respectively.}
\label{fig:radialdrift}
\end{figure}

To get an idea about how severe is the migration of pulsars along the radial
direction, in Figure \ref{fig:radialdrift} I have plotted the distribution of
the initial galactocentric radius of those pulsars which end up in the Solar
neighbourhood ($R\sim 6-10$ kpc) after their evolution. We can see that
the migration becomes very significant even in 10 Myr, for velocities of 250
km/sec. At 100 Myr, practically the objects come from everywhere to the solar
neighbourhood. This suggests that a complete understanding of the kinematical
properties of pulsars is possible only if we study their evolution in the full
galactic potential, and not in the approximated `local potential'.

\section{Velocities of millisecond pulsars}
The population of millisecond pulsars (MSPs) differ from the ordinary pulsars in
many properties. First of all, they are much older (more than $10^9$ years
old). Moreover, a good fraction of them are in binary systems. These make the
kinematic properties of millisecond pulsars quite distinctly different from the
ordinary population.

\begin{figure}
\epsfig{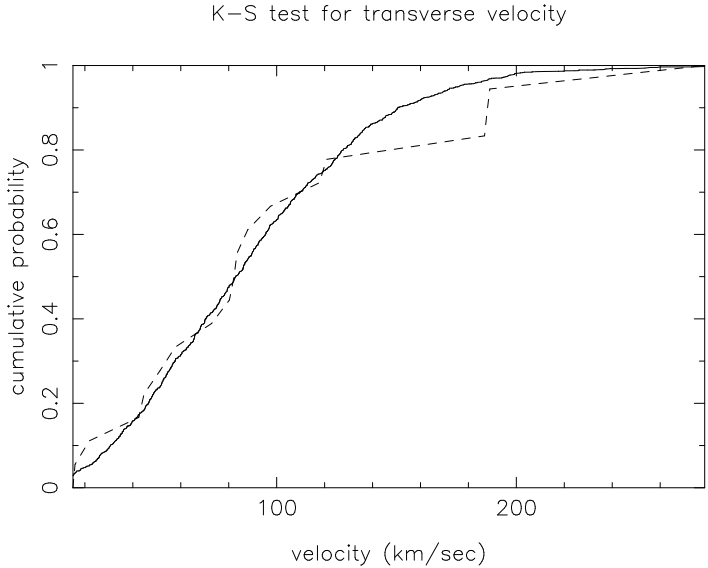}
\caption[]{Cumulative distribution of the observed transverse velocities of
millisecond pulsars (`dash' line). Solid line is the result of the simulation,
with a Maxwellian speed distribution of 1--D r.m.s. of 100 km/sec, and maximum
age of about $10^9$ years. The K--S probability of this fit is 61\%. See text
for details.}
\label{fig:msvel}
\end{figure}

In the standard evolutionary scenario of MSPs (see Bhattacharya \& van den
Heuvel 1991 for a detailed review), they originate from Low Mass X-ray Binaries
(LMXBs). These LMXBs acquire their space velocities during the primary (and the
only) Supernova explosion in the system. Even if the supernova explosion is a
symmetric one, the surviving binaries will acquire some space velocity, since
the explosion is symmetric only with respect to the exploding star, and not with
respect to the centre of mass of the binary system. During this process, a
significant fraction of the binary systems may get disrupted. The survival
probability gets reduced further if the explosion is an asymmetric one (Flannery
\& van den heuvel 1975; Hills 1983; Ramachandran \& Bhattacharya 1997; Tauris \&
Bailes 1996).

\begin{table}
\begin{center}
\begin{tabular}{lcccccccc}\hline
{\bf Jname}  & {\bf l} & {\bf b} & {\bf $\mu_{\alpha}$} & {\bf err} & 
{\bf $\mu_{\delta}$} & {\bf err} & {\bf d} & {\bf $V_t$} \\ 
  & (deg) & (deg) & (mas/yr) &  & (mas/yr) &  & (km/s) \\ \hline\hline
J0437-4715 &  253.394 & -41.964 &   114     & 2     & -72    &  4    &   0.14 &    88 \\
J0613-0200 &  210.41  &  -9.30  &     2.0   & 0.4   &  -7    &  1    &   2.19 &    58 \\
J0711-6830 &  279.53  & -23.28  &   -15.7   & 0.5   &  15.3  &  0.6  &   1.03 &   118 \\
J1024-0719 &  251.70  &  40.52  &   -41     & 2     & -70    &  3    &   0.35 &   121 \\
J1045-4509 &  280.85  &  12.25  &    -5     & 2.0   &  6     & 1     &  3.24  &  188 \\
J1300+1240 &  311.301 &  75.414 &    46.4   & 0     & -82.9  &  0    &   0.62 &   279 \\
J1455-3330 &  330.72  &  22.56  &     5     & 6.0   &  24    & 12    &   0.74 &    97 \\
J1603-7202 &  316.63  & -14.50  &    -3.5   & 0.3   &  -7.8  &  0.5  &   1.64 &    73 \\
J1643-1224 &    5.67  &  21.22  &     3.0   & 1.0   &  -8.0  &  5    &  $>$4.86 &    Undef \\
J1713+0747 &   28.751 &  25.223 &     4.9   & 0.3   &  -4.1  &  1.0  &   0.89 &    22 \\
J1730-2304 &    3.14  &   6.02  &    20.5   & 0.4   &   0.0  &  0    &   0.51 &    51 \\
J1744-1134 &   14.79  &   9.18  &    18.64  & 0.08  & -10.3  &  0.5  &   0.17 &    15 \\
J1857+0943 &   42.290 &   3.061 &    -2.94  & 0.04  &  -5.41 &  0.06 &   1.00 &    16 \\
J1911-1114 &   25.14  &  -9.58  &    -6     & 4     &  23    & 13    &   1.59 &   187 \\
B1937+21   &   57.509 &  -0.290 &    -0.130 & 0.008 & -0.464 & 0.009 &  3.58  &   80 \\
B1957+20   &   59.197 &  -4.697 &   -16.0   & 0.5   & -25.8  &  0.6  &   1.53 &   189 \\
J2019+2425 &   64.746 &  -6.624 &    -9.9   & 0.7   & -21.3  &  1.4  &   0.91 &    83 \\
J2124-3358 &   10.93  & -45.44  &   -14     & 1     & -47    &  1    &   0.24 &    45 \\
J2129-5721 &  338.01  & -43.57  &     7     & 2     & -4     & 3     & $>$2.55  &   Undef \\
J2145-0750 &   47.78  & -42.08  &    -9.1   & 0.7   & -15    &  2    &   0.50 &    43 \\
J2322+2057 &   96.515 & -37.31  &   -17     & 2     & -18    &  3    &   0.78 &
82.24 \\ \hline
\end{tabular}
\caption[]{List of measured proper motions of millisecond pulsars. Columns 2 \&
3 give the galactic longitude and latitude, 4--7 give the proper motions along
RA and Dec and their errors, columns 8 \& 9 give the distance calculated using
the Taylor \& Cordes (1993) model, and transverse velocity, respectively. For
those pulsars for which we know only a lower limit to distance, velocity has
been specified as ``Undef'' (undefined).}
\label{tab:mspsr}
\end{center}
\end{table}

Over past few years, many detailed analyses have been done to understand the
kinematic properties of these two populations. Through their analysis, van
Paradijs and White (1995) argued that to explain the distribution of LMXBs we
need high asymmetric kick velocities, and the distribution is consistent with
the kick speed distribution given by Lyne \& Lorimer (1994). Through their Monte
Carlo simulation, Brandt \& Podsiodlowski (1995) supported this
conclusion. Tauris \& Bailes (1996), from their evolutionary calculations,
showed that it is almost impossible to produce any MSP system with speeds $\gsim
250$ km/sec. Later, Ramachandran \& Bhattacharya (1997) showed through their
detailed Monte Carlo analysis of the evolution of LMXBs and MSPs in the Galactic
potential, that the spatial distribution of LMXBs and the speed distribution of
MSPs are consistent with the Lyne \& Lorimer distribution, but it is even more
consistent with even `zero' kick speed! This uncertainty (mainly by low number
statistics) was again shown by the analysis by Kalogera (1998) where she showed
that the distribution is consistent with speeds in the range 100 - 500 km/sec.

Given all these, let us see what we can infer from our simple simulation with
new proper motion measurements of many MSPs. Table \ref{tab:mspsr} gives a list
of MSPs with their measured proper motions and transverse velocities. It is
understandable that these objects, on the average, are moving slower than the
ordinary pulsars, since high velocities acquired during the supernova explosion
would have disrupted the binary system.

In order to get an idea about their intrinsic birth speeds, I have evolved the
sample pulsars for about $2\times 10^9$ years. Then, I have compensated for all
the observational selection effects. For this process, I have assumed that the
Parkes 70 cm survey (Manchester et al. 1996) can observe all directions in the
Galaxy. This is just to improve the statistics in the final `observable'
distribution of pulsars in the simulation. Then, from the velocities of those
`observable' pulsars, I have calculated their transverse velocity as measured
from the position of the Sun, so that they can be directly compared with the
measured transverse velocities of MSPs. Figure \ref{fig:msvel} shows the
cumulative distribution of the `true' distribution (`dash' line) and the
simulated distribution for an assumed initial Maxwellian speed distribution with
1-D r.m.s. of 100 km/sec (solid line). The Kolmogorov-Smirnov probability (K-S
probability) of this fit comes to about 61\%. The K-S probability of many other
distributions seem to be less than this value. For instance, for $\sigma_v=50$
km/sec, it is 20\%. And for $\sigma_v$ of 250 and 500 km/sec, the probabilities
were 12\% and 0\%, respectively.

\section{Observational evidences pertinent to kick mechanisms}
From various reasons we presented above, it is clear that the formation process
of neutron stars must be asymmetric, so that pulsars gain significant
velocities. The reasons, as we saw, are purely empirical, with different kinds
of observations pointing to the existance of an impulsive transfer of momentum
to the protoneutron star at birth (Shklowskii 1970; Gunn \& Ostriker 1970; van
den Heuvel \& van Paradijs 1997). The mechanisms suggested for this range from
hydrodynamical instabilities to those in which asymmetric neutrino emission is
postulated (Burrows 1987; Keil et al. 1996; Horowitz \& Li 1997; Lai \& Qian
1998; Spruit \& Phinney 1998). 

Whatever is the mechanism to create such an asymmetry, it is important to
understand if the direction of asymmetry is random, or it is associated with
some physical property of the system. Two such axes are the rotation and the
dipole magnetic axes of the neutron star. Many mechanisms suggested to produce
the asymmetry have invoked both these axes for the direction of the asymmetry
(Harrison \& Tademaru 1975a,b; Burrows \& Hayes 1996; Kusenko \& Segre 1996). 

The recent analysis by Deshpande, Ramachandran \& Radhakrishnan (1999) explores
into the observational tests to prove (or disprove) the mechanisms predicting
any relation between the direction and magnitude of the observed velocities and
the magnetic and the rotation axis of the star. As they show, observations do
not support any relation between ($i$) the magnitudes of velocities and magnetic
field, and ($ii$) the direction of velocities and the magnetic axis and 
rotation axis.

The recent work by Spruit \& Phinney (1998) suggests that the cores of the
progenitors of neutron stars cannot have the angular momentum to explain the
rotation of pulsars. They propose that the rotation of pulsars and their spatial
velocities must have a common origin. This suggestion was first made by Burrows
et al. (1995). As Spruit \& Phinney state, if the asymmetry is not directed
radially during the formation, then the star acquires both linear and angular
momentum. Cowsik (1998) has also advanced such a possibility. 

\begin{figure}
\epsfig{file=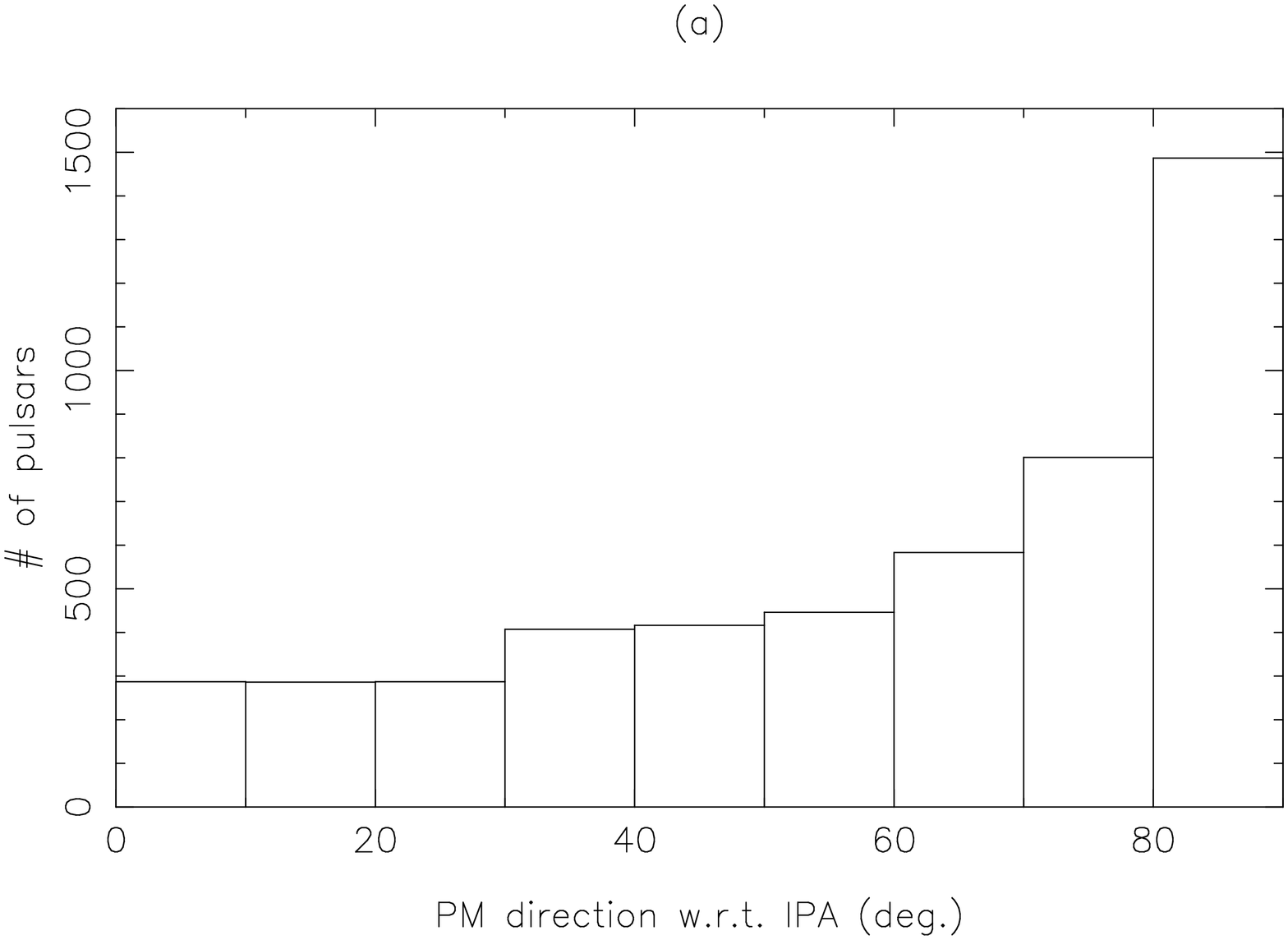,width=6.5cm,angle=-90}
\epsfig{file=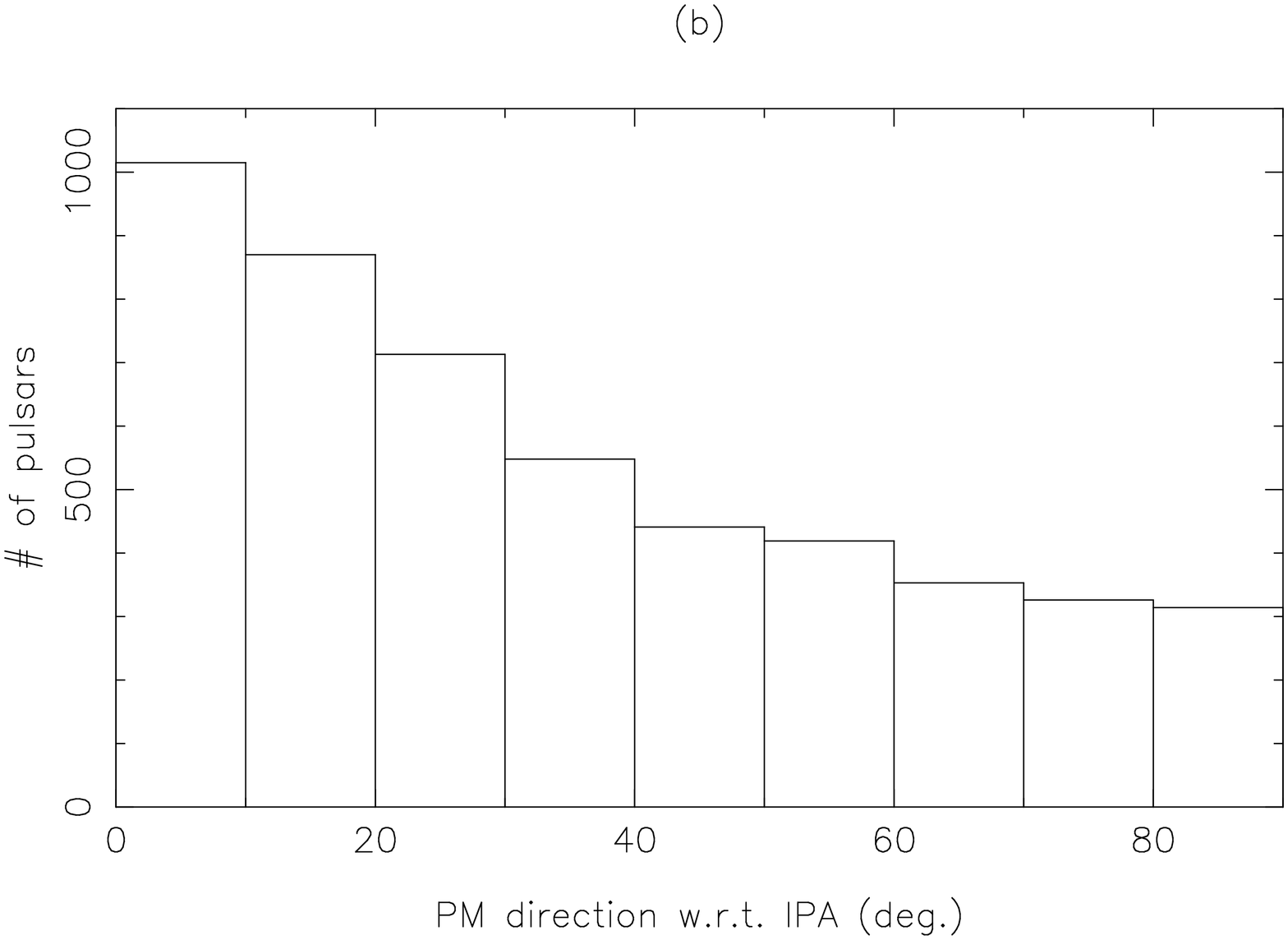,width=6.5cm,angle=-90}
\caption{Distribution of angles between the proper motion and the spin axis
projected in the plane of the sky (Deshpande, Ramachandran \& Radhakrishnan
1999). ({\bf a}) For a single non-radial impulse of any duration. The
distribution is biased towards 90 degrees, ({\bf b}) for a number of
long-duration impulses. The bias is towards zero degrees.}
\label{fig:angdiff}
\end{figure}

In order to test this hypothesis, we need to understand a number of
possibilities. In this case, the variables are, ($i$) number of impulses which
the (proto-) neutron star receives, ($ii$) duration of each impulse, ($iii$)
direction of each impulse. Let us assume that the direction of the impulses are
random. From the numerical simulations of Deshpande et al. (1999), it is clear
that if we have only one impulse (of any duration), the resultant direction of
the velocity is perpendicular to the direction of the spin axis. After taking 
account of the projection effects in the plane of the sky, they show that the angle
between the rotation and the proper motion axes should be biased towards
$90^{\circ}$. However, this is {\it not} what is seen observationally. Figure
\ref{fig:angdiff}a shows that the distribution is consistent with a uniform
distribution from zero to 90 degrees. Therefore, single impulses being
responsible for both the velocity and the rotation of the star can be ruled out.

If the number of impulses are more, then the problem becomes complicated, and is
a sensitive function of the duration of each of the kicks. First of all, we
expect that the strength of the impulses required to produce the observed range
of velocities goes down by $\sqrt{N}$, where $N$ is the number of impulses. The
velocity dispersion as a function of the duration of a momentum impulse of a
given magnitude remains constant up to a certain critical duration ($\tau_c$),
above which the azimuthal averaging reduces the resultant velocities. For
impulse durations much smaller than $\tau_c$, both the linear and angular
momenta grow as $\sqrt{N}$, but the angle between them becomes random. However,
for relatively longer duration impulses a significant preference of the
direction of the linear momentum develops towards the spin axis, which itself is
evolving. This is shown in Figure \ref{fig:angdiff}b, where the angle has a
preferential bias towards zero degrees. Since observations do not support this
either, we can conclude that long duration multiple impulses can also be ruled
out.

To summarise, observationally we can conclude the following: (1) velocity
magnitudes are {\it not} correlated to the magnetic field strength, (2) pulsar
velocities are {\it not} in the direction of their magnetic axis or rotation
axis, (3) if non-radial kicks produce both pulsar velocities and rotation, then
the single kicks of {\it any duration} are ruled out, and (4) multiple kicks of
long duration are ruled out.

\section{Concluding remarks}
I have given a brief summary of the kinematic properties of pulsars in this
article. Though there are some pulsars which have significantly large spatial
velocities (of the order of 1000 km/sec), on the average, there are strong
evidences to show that the population has a three dimensional speed of about 200
to 250 km/sec. All the derived speed distributions show that the predominant
number of pulsars move with such speeds. There are many empirical evidences to
show that pulsars acquire a significant fraction of their velocities through an
asymmetric supernova explosion.

Pulsar population shows a great deal of migration along both galactocentric
radius and height from the plane. Because of this, they show significant
asymmetric drift. This is one of the remarkable properties observed on pulsar
population. 

Millisecond pulsars, due to their evolutionary history, have significantly lower
space velocities. Their average velocity is around 100 km/sec, and their
kinematic properties in the Galaxy match well with those of the Low Mass X-Ray
Binaries, supporting the idea that MSPs are born from LMXBs.

There are many models which have advanced many different ways of producing
asymmetries during supernova explosions. Given the latest high-quality proper
motion measurements and polarisation information with which we can determine the
direction of rotation axis in the plane of the sky, one can test these
models. These tests reveal that
\begin{itemize}
\item Mechanisms predicting correlation between the rotation axis and the pulsar
      velocities are ruled out.
\item There is no correlation between the direction of magnetic axis and the
      direction of velocities.
\item There is no correlation between the strength of the magnetic fields and the
      magnitude of velocities.
\item If non-radial asymmetric impulses is responsible for both velocity and
      rotation of pulsars, then the scenario where the star receives a single
      impulse (of any duration) can be ruled out.
\item Even multiple impulses of long durations can be ruled out.
\end{itemize}
In principle, multiple short-duration impulses are not ruled out, as they would
produce randomly oriented rotation and velocity axis. Similarly, we cannot rule
out the possibility of having completely radial kicks as the origin of pulsar
velocities. In this case, then the rotation of the pulsar will have nothing to
do with the asymmetric explosion.

\section*{Acknowledgements}
I would like to thank D. Bhattacharya, A. Blaauw, A. A. Deshpande,
V. Radhakrishnan and E. P. J. van den Heuvel, for many fruitful discussions.


\begin{thebibliography}{}
\bibitem[]{} Anderson, B. and Lyne, A. G. (1983) {\it Nat}, {\bf 303}, 597
\bibitem[]{} Bagot, P. (1996) {\it A\&A}, {\bf 314}, 576
\bibitem[]{} Bhattacharya, D. and van den Heuvel, E. P. J. (1991) {\it
           Phys. Rep.}, {\bf 203}, 1
\bibitem[]{} Blaauw, A. (1961), {\it Bull. Astr. Inst. Netherlands}, {\bf 15}, 265
\bibitem[]{} Blaauw, A. and Ramachandran, R. (1998), {\it JApA}, {\bf 19}, 19
\bibitem[]{} Brandt, W. N. and Podsiadlowski, Ph. (1995), {\it MNRAS}, {\bf 277}, 35
\bibitem[]{} Burrows, A. (1987) {\it ApJ}, {\bf 318}, L57
\bibitem[]{} Burrows, A. and Hayes, J. (1996) {\it PRL}, {\bf 76}, 352
\bibitem[]{} Burrows, A., Hayes, J. and Fryxell, B. A. (1995) {\it ApJ}, {\bf 450}, 830
\bibitem[]{} Cordes, J. M. and Chernoff, D. F. (1997) {\it ApJ}, {\bf 482}, 971
\bibitem[]{} Cowsik, R. (1998) {\it A\&A}, {\bf 340}, L65
\bibitem[]{} Deshpande, A. A. and Ramachandran, R. (1998) {\it MNRAS}, {\bf 300}, 577
\bibitem[]{} Deshpande, A. A., Ramachandran, R. and Radhakrishnan, V. (1999) {\it A\&A},
           Accepted for publication
\bibitem[]{} Flannery, B. P. and van den Heuvel, E. P. J. (1975)  {\it A\&A}, {\bf 39}, 61
\bibitem[]{} Fryer, C. and Kalogera, V. (1997) {\it ApJ}, {\bf 489}, 244
\bibitem[]{} Gunn, J. E. and Ostriker, J. P. (1970) {\it ApJ}, {\bf 160}, 979
\bibitem[]{} Hansen, B. M. S. and Phinney, E. S. (1997) {\it MNRAS}, {\bf 291}, 569
\bibitem[]{} Harrison, E. R. and Tademaru, E. (1975a) {\it ApJ}, {\bf 201}, 447
\bibitem[]{} Harrison, E. R. and Tademaru, E. (1975b) {\it Nat}, {\bf 254}, 676
\bibitem[]{} Hills, J. G. (1983) {\it ApJ}, {\bf 267}, 322
\bibitem[]{} Horowitz, C. J. and Li, G. (1998), {\it PRL}, {\bf 80}, 3694
\bibitem[]{} Kalogera, V. (1996) {\it ApJ}, {\bf 471}, 352
\bibitem[]{} Kalogera, V. (1998) in {\it Neutron stars and Pulsars} held in
           Rikkoyo University, Japan. eds. N. Shibazaki et al. p.27
\bibitem[]{} Kaspi, V. M. Bailes, M., Manchester, R. N. et al. (1996) {\it Nat},
           {\bf 381}, 584
\bibitem[]{} Keil, W., Janka, H.-Th. and Muller, E. (1996) {\it ApJ}, {\bf 473}, L111
\bibitem[]{} Kuijken, K. and Gilmore, G. (1989) {\it MNRAS}, {\bf 239},571
\bibitem[]{} Kusenko A. and Segre, G. (1996) {\it PR}L, {\bf 77}, 24
\bibitem[]{} Lai, D. (1996) {\it ApJ}, {\bf 466}, 35
\bibitem[]{} Lai, D. and QIan, Y. (1998) {\it ApJ}, {\bf 495}, 103
\bibitem[]{} Lipunov, V. M., Postnov, K. A. and Prokhorov, M. E. (1996) {\it
           A\&A}, {\bf 310}, 489
\bibitem[]{} Lorimer, D. R., Lyne, A. G. and Anderson, B. (1995) {\it MNRAS}, {\bf
           275}, L16
\bibitem[]{} Lyne, A. G. and Lorimer, D. R. (1994) {\it Nat}, {\bf 369}, 127
\bibitem[]{} Manchester, R. N., Lyne, A. G., D'Amico, N. et al. (1996) {\it
           MNRAS}, {\bf 279}, 1235
\bibitem[]{} Mihalas, D. and Binney, J. (1981) {\it Galactic Astronomy},
           W. H. Freeman, New York
\bibitem[]{} Morris, D., Radhakrishnan, VB. and Shukre, C. S. (1976) {\it Nat},
           {\bf 260}, 124
\bibitem[]{} Narayan, R., Piran, T. and Shemi, A. (1991) {\it ApJ}, {\bf 379}, 17
\bibitem[]{} Phinney, E. S. (1991) {\it ApJ}, {\bf 380}, 17
\bibitem[]{} Portegies Zwart, S. F. and Spreeuw, H. N. (1996) {\it A\&A}, {\bf 312}, 670
\bibitem[]{} Ramachandran, R. (1996) {\it Ph.D. Thesis}, Osmania University
\bibitem[]{} Ramachandran, R. and Bhattacharya, D. (1997) {\it MNRAS}, {\bf 288}, 565
\bibitem[]{} Shklowskii, I. S, (1970) {\it Astr. Zu.}, {\bf 46}, 715
\bibitem[]{} Spruit, H. C. and Phinney, E. S. (1998) {\it Nat}, {\bf 393}, 139
\bibitem[]{} Tauris, T. and Bailes, M. (1996) {\it A\&A}, {\bf 315}, 432
\bibitem[]{} Taylor, J. H. and Cordes J. M. (1993), {\it ApJ}, {\bf 411},674
\bibitem[]{} Trimble, V. and Rees, M. (1971) {\it IAU Symposium no. 46} held at
           Jodrell Bank, 1970, eds. Rodney Deane Davies, Francis Graham-Smith,
           Dordrecht, Reidel, p.273
\bibitem[]{} van den Heuvel, E. P. J. and Lorimer, D. R. (1996) {\it MNRAS}, {\bf 283}, 37
\bibitem[]{} van den Heuvel, E. P. J. and van Paradijs, J. (1997) {\it ApJ}, {\bf 483}, 399
\bibitem[]{} van Paradijs, J. and White, N. (1995), {\it ApJ}, {\bf 447}, 33
\bibitem[]{} Wex, N., Kalogera, V. and Kramer, M. (2000) {\it ApJ}, {\bf 528} (in press,
           astro-ph/9905331)
\end{thebibliography}
\end{document}